\documentclass[twocolumn, amssymb, nobibnotes, aps, groupedaddress, prl, reprint, superscriptaddress, showpacs]{revtex4-1}

\usepackage{natbib}
\usepackage{graphicx}
\usepackage{amsmath}
\usepackage{amssymb}
\usepackage[caption=false]{subfig}
\usepackage{array}
\newcolumntype{C}[1]{>{\centering\let\newline\\\arraybackslash\hspace{0pt}}m{#1}}
\newcolumntype{L}[1]{>{\raggedright\let\newline\\\arraybackslash\hspace{0pt}}m{#1}}
\newcolumntype{R}[1]{>{\raggedleft\let\newline\\\arraybackslash\hspace{0pt}}m{#1}}
\begin{document}

\title{Improved Determination of the Neutron Lifetime}%

\author{A.T.~Yue}
\email{andrew.yue@nist.gov}
\affiliation{Institute for Research in Electronics and Applied Physics, University of Maryland, College Park, Maryland 20742}
\affiliation{National Institute of Standards and Technology, Gaithersburg, Maryland 20899}
\affiliation{University of Tennessee, Knoxville, Tennessee 37996}
\author{M.S.~Dewey}
\affiliation{National Institute of Standards and Technology, Gaithersburg, Maryland 20899}
\author{D.M.~Gilliam}
\affiliation{National Institute of Standards and Technology, Gaithersburg, Maryland 20899}
\author{G.L.~Greene}
\affiliation{University of Tennessee, Knoxville, Tennessee 37996}
\affiliation{Oak Ridge National Laboratory, Oak Ridge, Tennessee 37831}
\author{A.B.~Laptev}
\affiliation{Tulane University, New Orleans, Louisiana 70118}
\affiliation{Los Alamos National Laboratory, Los Alamos, New Mexico 87545}
\author{J.S.~Nico}
\affiliation{National Institute of Standards and Technology, Gaithersburg, Maryland 20899}
\author{W.M.~Snow}
\affiliation{Indiana University, Bloomington, Indiana 47408}
\author{F.E.~Wietfeldt}
\affiliation{Tulane University, New Orleans, Louisiana 70118}

\begin{abstract}
The most precise determination of the neutron lifetime using the beam method was completed in 2005 and reported a result of $\tau_{n} = \left(886.3 \pm 1.2\left[\textrm{stat}\right] \pm 3.2\left[\textrm{syst}\right]\right)$ s.  The dominant uncertainties were attributed to the absolute determination of the fluence of the neutron beam (2.7 s).  The fluence was measured with a neutron monitor that counted the neutron-induced charged particles from absorption in a thin, well-characterized ${}^{6}$Li deposit.  The detection efficiency of the monitor was calculated from the areal density of the deposit, the detector solid angle, and the evaluated nuclear data file, ENDF/B-VI ${}^{6}$Li(n,t)${}^{4}$He thermal neutron cross section.  In the current work, we have measured the detection efficiency of the same monitor used in the neutron lifetime measurement with a second, totally-absorbing neutron detector. This direct approach does not rely on the ${}^{6}$Li(n,t)${}^{4}$He cross section or any other nuclear data. The detection efficiency is consistent with the value used in 2005 but was measured with a precision of 0.057 \,\%, which represents a fivefold improvement in the uncertainty.  We have verified the temporal stability of the neutron monitor through ancillary measurements, allowing us to apply the measured neutron monitor efficiency to the lifetime result from the 2005 experiment.  The updated lifetime is $\tau_{n} = \left(887.7 \pm 1.2\left[\textrm{stat}\right] \pm 1.9\left[\textrm{syst}\right]\right)$ s.
\end{abstract}
\maketitle

The accurate determination of the mean lifetime of the free neutron addresses fundamentally important questions in particle physics, astrophysics, and cosmology \cite{Dubbers11,Wietfeldt11}.  To date, two distinct experimental strategies have been used to accurately measure the neutron lifetime.  In the first, or beam method, the rate of neutron decay $dN/dt$ and the number of neutrons $N$ in a well-defined volume of a neutron beam are determined.  The neutron lifetime is determined from the differential form of the exponential decay function $dN/dt = -N/{\tau_{n}}$.  In the second, or bottle method, neutrons of sufficiently low energy are confined in a trap or bottle established by some combination of material walls, magnetic fields, and/or gravity.  The number of neutrons in the bottle at various times $t$ is measured and fit to the exponential decay function $N(t) = N(0)e^{-t/{\tau_{n}}}$ in order to extract $\tau_{n}$.

\begin{figure}[htbp]
  \centering
  \includegraphics[width=0.49\textwidth]{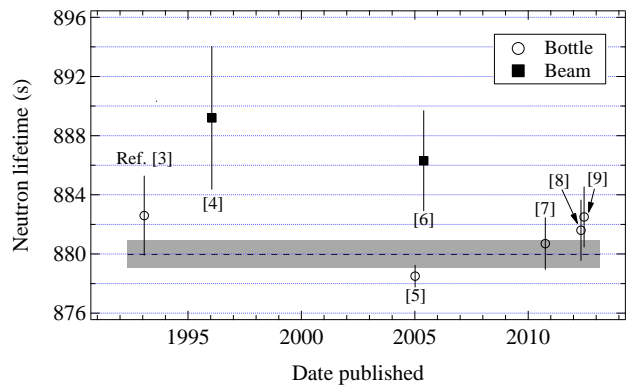}
  \caption{(Color online) The neutron lifetime measurements used in the 2013 PDG world average.  The weighted mean and 1 $\sigma$ uncertainty (inflated by scale factor $\sqrt{\chi^{2}/\textrm{d.o.f.}} = 1.53$, following PDG procedures) of the data set is represented by the dashed line and shaded band.}
  \label{fig:tnplot}
\end{figure}

Measurements used to form the 2013 Particle Data Group (PDG) world average value for $\tau_{n}$ include the five bottle and two beam measurements shown in Fig.~\ref{fig:tnplot} \cite{PDG12}.  While there is currently reasonable internal consistency among the bottle and among the beam determinations, the two sets differ from each other by 2.6 $\sigma$ (where $\sigma$ is one standard deviation).  Historical discrepancies among independent bottle experiments and between bottle and beam measurements suggest that it is highly desirable to not only improve the experimental limits on $\tau_{n}$ but to also carefully study systematic effects in all methods.  We have completed an investigation into the dominant systematic uncertainty in the most precise beam neutron lifetime measurement, resulting in confirmation of the accuracy of the fluence measurement technique and a reduction in the total uncertainty in the lifetime result.

The beam neutron lifetime measurement with the lowest quoted uncertainty was performed at the National Institute of Standards and Technology (NIST) Center for Neutron Research (NCNR) in Gaithersburg, MD \cite{Dewey03,Nico05}.  The experiment used an electromagnetic proton trap to determine the absolute rate of neutron decay with known absolute efficiency in a well-defined volume of the beam and a neutron fluence monitor to measure the absolute density of neutrons in the proton trap.  A detailed description of the decay detection method is given in reference \cite{Nico05}.

The neutron density in the proton trap was determined by measuring the rate of charged particle products from the ${}^{6}$Li(n,t)${}^{4}$He reaction.  The neutron beam passed through a thin deposit of ${}^{6}$LiF and the rate of reaction products was measured by surface barrier detectors masked by precision apertures.  Ignoring backscattering and below-energy threshold events ($<{10}^{-4}$ of all incident charged particles), incident alphas and tritons are detected with unit efficiency, and thus the detection efficiency for charged particles is given by the solid angle subtended by the apertures.  The solid angle was measured to a precision of 0.1 \,\% by both contacting metrology and, independently, using a calibrated $\alpha$-source.  The geometry of the detector array is such that the solid angle is first-order insensitive to shifts in source position.  The amount of ${}^{6}$Li evaporated on the target deposit was determined by comparison of thermal neutron-induced charged particle reaction rates to two deposits of the same nominal mass \cite{Scott95}.  These two comparison deposits were later destructively analyzed by isotope dilution mass spectrometry \cite{Lamberty91}, establishing the neutron-induced activity per ${}^{6}$Li atom in the deposit.  The distribution of ${}^{6}$Li on the deposit was calculated from the geometry of the evaporator and verified through a combination of mechanical and optical techniques.  The sharpness of the deposit edge was measured by Talystep profilometer and the deposit diameter was measured by traveling microscope and Abbe comparator \cite{Pauwels95}.  These measurements were used to determine an average areal density $\bar{\rho} = (39.30 \pm 0.10)$ $\mu$g/cm${}^{2}$, where the symbol $\pm$ here and throughout the text corresponds to the standard (1 $\sigma$) uncertainty.  The ${}^{6}$Li thermal neutron cross section used in the 2005 measurement is ($941.0 \pm 1.3$) b and comes from the evaluated nuclear data file ENDF/B-VI \cite{ENDF6}.  The total uncertainty in the determination of the average neutron density was approximately 0.3 \,\%.  This uncertainty dominates the reported uncertainty on $\tau_{n}$ \cite{Dewey03,Nico05}.  We note the other beam lifetime measurement quoted in the PDG performed the neutron density determination in similar fashion \cite{Byrne90,Byrne96}.

The quoted ${}^{6}$Li(n,t)${}^{4}$He cross section represents a best fit to a very large set of nuclear data parameters.  A new evaluation (ENDF/B-VII.0) was released shortly after publication of the 2005 lifetime result \cite{ENDF7}.  The updated ${}^{6}$Li(n,t)${}^{4}$He cross section is $\sigma_{0} = (938.5 \pm 1.3)$ b, which is in slight disagreement with the ENDF/B-VI value.  This unsatisfactory situation requires that the neutron lifetime result be adjusted each time a new evaluation of the ${}^{6}$Li(n,t)${}^{4}$He cross section is adopted.  In order to avoid this, and eliminate all systematic effects associated with the evaluated cross section, we have performed a direct, first-principles measurement of the neutron monitor efficiency.  This measurement serves several purposes:
\begin{itemize}
\item{It can be used to improve the neutron fluence determination in the 2005 lifetime result;}
\item{It reduces the overall uncertainty on the 2005 lifetime result;}
\item{It removes the necessity to alter the 2005 lifetime result each time there is a change in the accepted value of the ${}^{6}$Li(n,t)${}^{4}$He cross section;}
\item{When combined with proton detection improvements learned in a similar experiment \cite{Cooper10}, it improves the achievable uncertainty in a future run of the neutron lifetime experiment to approximately 1 s \cite{Dewey09}.}
\end{itemize}

\begin{figure}[htbp]
  \centering
  \includegraphics[width=0.49\textwidth]{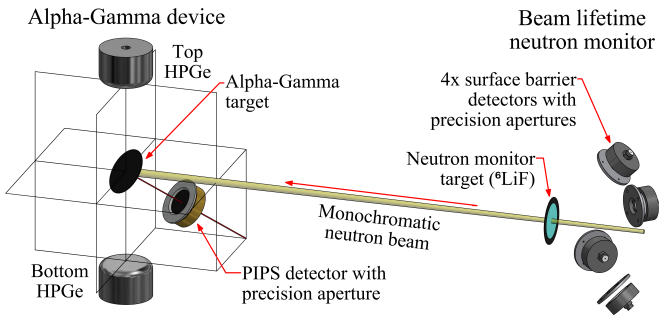}  
  \caption{(Color online) Neutron monitor efficiency measurement setup and Alpha-Gamma detection geometry.}
  \label{fig:AGFMDetGeo}
\end{figure}

Direct measurement of the neutron monitor efficiency is accomplished by operating the neutron monitor on a monochromatic beam with total neutron rate $R_{n}$ and wavelength $\lambda_{\textrm{mono}}$. The observed rate of alphas and tritons ($r_{\alpha,t}$) in the monitor is
\begin{equation}
\label{eqn:rat}
r_{{\alpha},t} = \epsilon_{0}\frac{\lambda_{\textrm{mono}}}{\lambda_{0}}R_{n},
\end{equation}
where $\epsilon_{0}$ is the detection efficiency of the neutron monitor for a thermal neutron (wavelength $\lambda_{0} = 0.1798$ nm).  By running the neutron monitor with a device that directly measures $R_{n}$ and a device that measures $\lambda_{\textrm{mono}}$, one can determine $\epsilon_{0}$ without reference to the ${}^{6}$Li thermal neutron cross section or the amount of ${}^{6}$Li present in the target.  The total neutron rate is measured in the Alpha-Gamma device, which is an absolute counter for cold and thermal neutrons based on the counting of prompt gamma rays from a totally absorbing, thick ${}^{10}$B${}_{4}$C target.  As shown in Fig. \ref{fig:AGFMDetGeo}, the ${}^{10}$B${}_{4}$C target is positioned to face a passivated implanted planar silicon (PIPS) detector, and is viewed from above and below by high-purity germanium (HPGe) detectors.  The gamma detection efficiency is determined in a calibration procedure in which the well known activity of a ${}^{239}$Pu $\alpha$-source is successively transferred through a series of intermediate steps \cite{Gilliam89}.

The neutron monitor efficiency measurement was completed in 2011 on beam line NG-6m at the NCNR \cite{Yue11,AGInstrument13}.  The measurement was performed at a beam wavelength of $\lambda_{\textrm{mono}} = (0.49605 \pm 0.00012)$ nm with a measured detection efficiency of $\epsilon^{\textrm{meas}} = (8.5797 \pm 0.0048) \times {10}^{-5}$.  The thermal neutron detection efficiency is
\begin{equation}
\epsilon_{0}^{\textrm{meas}} = \epsilon^{\textrm{meas}}\frac{\lambda_{\textrm{mono}}}{\lambda_{0}} = (3.1098 \pm 0.0017) \times{10}^{-5}.
\end{equation}
This is in agreement with the calculated value of $\epsilon_{0}$ used in the 2005 lifetime result
\begin{equation}
\label{eqn:e0calc}
\begin{split}
\epsilon^{\textrm{calc}}_{0} & = 2\frac{N_{A}}{A}\Omega\left(0,0\right)\rho\left(0,0\right)\sigma_{0}\\
 &  =  (3.1148 \pm 0.0094) \times{10}^{-5},
\end{split}
\end{equation}
where $N_{A}$ is the Avogadro constant, $A$ is the atomic weight of ${}^{6}$Li, $\Omega\left(0,0\right)$ is the detector solid angle to the center of the target, and $\rho\left(0,0\right)$ is the areal density at the center of the target.

The fivefold improvement in the determination of the neutron monitor efficiency significantly reduces the total uncertainty in the updated lifetime.  As noted previously, the neutron monitor that was measured in the process described above was the same one used to assess the neutron fluence in the 2005 lifetime experiment.  Nonetheless, it cannot be excluded that the efficiency of the neutron monitor has changed since the lifetime experiment was carried out.  The change would be attributable to a change in the solid angle subtended by the apertures ($\Delta\Omega$) and/or the areal density of the ${}^{6}$Li target deposit ($\Delta\rho$).  While we expect no change in these parameters, our ability to retroactively correct the 2005 lifetime result depends on the ability to verify that they have remained unchanged.

The solid angle subtended by the neutron monitor apertures was measured by an $\alpha$-source prior to the monitor efficiency measurement and was measured by $\alpha$-source and dimensional metrology after the completion of the monitor efficiency measurement.  The $\alpha$-source measured solid angle was $\Omega^{\alpha} = (4.2021 \pm 0.0014)\times{10}^{-3}$.  The solid angle as measured by dimensional metrology was $\Omega^{\textrm{DM}} = (4.2024 \pm 0.0010) \times{10}^{-3}$.  These results are in slight disagreement with the value of $\Omega = (4.196 \pm 0.004) \times{10}^{-3}$ used in the 2005 lifetime experiment.  In the new metrology campaign, the aperture diameters were determined by direct probe contact with the aperture cylinders at several bore depths.  In the previous campaign, the aperture diameters were determined from the intersection of the best-fit functions to measurements of the aperture backplanes and a precision tooling ball resting within the apertures.  The recent dimensional metrology results found that the aperture diameters were 0.06 \,\% larger than the diameters used in the 2005 lifetime experiment, which accounts for essentially all of the discrepancy between the two solid angle determinations.  A plausible explanation for the underasseessment of the aperture diameters in the previous campaign is that small ($\approx{10}^{-6}$ m) mechanical imperfections on the upper edge of the apertures kept the tooling ball from making contact with the apertures.  In addition to determining the aperture diameters to an order of magnitude higher precision, the new metrology result is not affected by the presence of small mechanical imperfections.  The discrepancy in $\Omega$ is attributed to measurement error in the previous, less precise dimensional metrology and $\alpha$-counting.  We conclude that $\Delta\Omega = 0$ and assign no additional systematic uncertainty to the lifetime.

A direct measurement of $\Delta\rho$ for the deposit used in the lifetime experiment requires destructive analysis.   This was not an acceptable option, so a non-destructive comparison technique was used to set bounds on $\Delta\rho$.  Fourteen ${}^{6}$Li deposits of three areal densities of approximately 20, 30, and 40 $\mu$g/cm${}^{2}$ were produced for the lifetime experiment.  The neutron induced activity for each deposit was measured on the same test beam with the same geometry \cite{Scott95}.  Three deposits (one of each density) were measured with the Alpha-Gamma device.  The measured detection efficiencies and the 1995 measurements of neutron induced activity (recorded in Table \ref{tab:BayesianData}) are proportional to the areal density of the deposits.  The two measurement sets can be related by a conversion constant $K$ and a density change $\Delta\rho$ in one of three models: 1) a surface loss model in which each deposit has lost an equivalent density $\Delta\rho$ that is proportional to the surface area of the deposit; 2) a handling loss model in which only the $39.30$ $\mu$g/cm${}^{2}$ deposit used in the lifetime experiment has lost density; 3) a no-loss model.  A Bayesian analysis \cite{Carlin95} was performed in which the observed data were simultaneously fit to the three models in order to determine the relative likelihood of each model.  The observables were assumed to be normally distributed and the prior distribution of $\Delta\rho$ was restricted to negative values (deposit density gain is unphysical).  At the 90 \,\% confidence level, the data favor the no loss model (3). For the less likely models (1 \& 2) that assumed some loss, the analysis found, on average, $\Delta\rho = 0.1$ \,\% of the lifetime foil density. We conclude that $\Delta\rho = 0$ but include, in our overall uncertainty budget, an uncertainty of 0.1 \,\% on $\Delta\rho$.

\begin{table}
\caption{\label{tab:BayesianData}Data used to determine limits on $\Delta\rho$.}
\centering
\begin{tabular}{C{2cm}C{3cm}C{2cm}}
\hline\hline\noalign{\smallskip}
Deposit areal density ($\mu$g/cm${}^{2}$) & Measured $\epsilon_{0}$ (${}\times{10}^{-5}$) & $r_{\alpha,t}$ from ref. \cite{Scott95} (s${}^{-1}$) \\
\noalign{\smallskip}\hline\noalign{\smallskip}
$19.94 \pm 0.05$ & $1.5731 \pm 0.0014$ & $600.83 \pm 0.33$\\
$28.58 \pm 0.07$ & $2.2689 \pm 0.0025$ & $862.23 \pm 0.68$\\
$39.30 \pm 0.10$ & $3.1098 \pm 0.0017$ & $1186.77 \pm 0.44$\\
\noalign{\smallskip}\hline\hline
\end{tabular}
\end{table}

The neutron lifetime is updated by applying the measured monitor efficiency and the two corrections for temporal drifts in the monitor components
\begin{equation}
\begin{split}
\tau_{n} & = \tau_{n}^{2005}\frac{\epsilon_{0}^{\textrm{calc}}}{\epsilon_{0}^{\textrm{meas}}}\left(1+\Delta\Omega\right)\left(1+\Delta\rho\right)\\
 & = (887.7 \pm 2.3) \textrm{ s}.
\end{split}
\end{equation}
The updated uncertainty budget for the neutron lifetime is presented in Table \ref{tab:2012tnbudget}.  The improved lifetime value is in agreement with the 2005 result.  Because the neutron fluence determination from the 2005 result and from this work are independent, a weighted average of the two could be performed to further reduce the uncertainty.  However, this small reduction would come at the cost of continued dependence on the ENDF-determined value for the ${}^{6}$Li(n,t)${}^{4}$He cross section.  As such, we do not adjust our uncertainty in this fashion and simply make use of the new neutron fluence determination.  Note that the updated neutron lifetime from this work does not include corrections for any proton counting systematic effects beyond those already addressed in Ref. \cite{Nico05}.

After replacing the 2005 lifetime result with the improved result, weighted fits to the beam and bottle lifetime results included in the 2013 PDG world average find that $\Delta\tau_{n} = \tau_{n}^{\textrm{beam}} - \tau_{n}^{\textrm{bottle}} = (888.0 \pm 2.1) \textrm{~s} - (879.6 \pm 0.8) \textrm{~s} = (8.4 \pm 2.2)$ s, a 3.8 $\sigma$ discrepancy.  It is important that this discrepancy be resolved with additional neutron lifetime measurements at increased precision using multiple techniques.  To that end, work has begun on improvements to the beam lifetime apparatus toward a goal of 1 s uncertainty in a new measurement.  While we believe the proton counting systematics to be accurate to the level of uncertainty quoted in Ref. \cite{Nico05}, further investigation into these systematic effects will be conducted as part of this new measurement \cite{Dewey09}.

\begin{table}
\caption{\label{tab:2012tnbudget}The new uncertainty budget for the neutron lifetime.  Corrections shown are relative to the 2005 beam lifetime result.}
\centering
\begin{tabular}{L{4cm}cc}
\hline\hline\noalign{\smallskip}
Source of uncertainty & Correction (s) & Uncertainty (s)\\
\noalign{\smallskip}\hline\noalign{\smallskip}
Improved neutron fluence determination & +1.4 & 0.5\\
Change in ${}^{6}$Li deposit mass & +0.0 & 0.9\\
\noalign{\smallskip}\hline\noalign{\smallskip}
Systematics unassociated with neutron fluence & & 1.7\\
\noalign{\smallskip}\hline\noalign{\smallskip}
Proton counting statistics & & 1.2\\
Neutron counting statistics & & 0.1\\
\noalign{\smallskip}\hline\noalign{\smallskip}
Total & +1.4 & 2.3\\
\noalign{\smallskip}\hline\hline
\end{tabular}
\end{table}
\begin{acknowledgments}
We thank W.F.~Guthrie for developing and performing the Bayesian analysis used to set limits on $\Delta\rho$.  We thank J.R.~Stoup for performing the dimensional metrology on the neutron monitor aperture rig.  We gratefully acknowledge the support of NIST (U.S. Department of Commerce), the U.S. Department of Energy Office of Nuclear Physics (Grants No. DE-SC0005925 and No. DE-FG02-03ER41258), and the National Science Foundation (Grants No. PHY-0855310, No. PHY-1068712, and No. PHY-1205266). W.M.S.~acknowledges support from the Indiana University Center for Spacetime Symmetries.
\end{acknowledgments}
\bibliography{tnshort}
\end{document}